\documentclass[aps,prb,twocolumn,superscriptaddress,showpacs,floatfix]{revtex4-1}
\usepackage{graphicx,hyperref}
\begin{document}

%\preprint{}

\title{Study of the optimal conditions for NV$^-$ center formation in type 1b diamond, using photoluminescence and positron annihilation spectroscopies.}


\author{J. Botsoa}
%\thanks{}
\affiliation{Conditions Extr\^emes et Mat\'eriaux : Haute Temp\'erature et Irradiation, CNRS UPR 3079, Orl\'eans, F-45071, France}
\affiliation{Laboratoire de Photonique Quantique et Mol\'eculaire, CNRS UMR 8537, \'Ecole Normale Sup\'erieure de Cachan, Cachan, F-94235, France}
\author{T. Sauvage}
\affiliation{Conditions Extr\^emes et Mat\'eriaux : Haute Temp\'erature et Irradiation, CNRS UPR 3079, Orl\'eans, F-45071, France}
\author{M.-P. Adam}
\affiliation{Laboratoire de Photonique Quantique et Mol\'eculaire, CNRS UMR 8537, \'Ecole Normale Sup\'erieure de Cachan, Cachan, F-94235, France}

\author{P. Desgardin}
\author{E. Leoni}
\author{B. Courtois} 
\affiliation{Conditions Extr\^emes et Mat\'eriaux : Haute Temp\'erature et Irradiation, CNRS UPR 3079, Orl\'eans, F-45071, France}
\author{F. Treussart}
\email{francois.treussart@ens-cachan.fr}
\affiliation{Laboratoire de Photonique Quantique et Mol\'eculaire, CNRS UMR 8537, \'Ecole Normale Sup\'erieure de Cachan, Cachan, F-94235, France}

\author{M.-F. Barthe}
\email{marie-france.barthe@cnrs-orleans.fr}
\affiliation{Conditions Extr\^emes et Mat\'eriaux : Haute Temp\'erature et Irradiation, CNRS UPR 3079, Orl\'eans, F-45071, France}


\date{\today}

\begin{abstract}
We studied the parameters to optimize the production of negatively-charged nitrogen-vacancy color centers (NV$^-$) in type~1b single crystal diamond using proton irradiation followed by thermal annealing under vacuum. Several samples were treated under different irradiation and annealing conditions and characterized by slow positron beam Doppler-broadening and photoluminescence (PL) spectroscopies. At high proton fluences another complex vacancy defect appears limiting the formation of NV$^-$. Concentrations as high as $2.3\times10^{18}$~cm$^{-3}$ of NV$^-$ have been estimated from PL measurements. Furthermore, we inferred the trapping coefficient of positrons by NV$^-$. This study brings insight into the production of a high concentration of NV$^-$ in diamond, which is of utmost importance in ultra-sensitive magnetometry and quantum hybrid systems applications.
\end{abstract}

%\keywords{}

\maketitle

\section{Introduction}
Owing to its unique properties of perfect photoluminescence stability and optically detectable electronic spin resonance with a long coherence time, the nitrogen-vacancy (NV$^-$) color center in diamond has been the object of a large number of studies over the past few years~\cite{Awschalom:ub,Jelezko:2011uz} from quantum information processing, to highly sensitive magnetometry~\cite{Balasubramanian:2008ga,Taylor:2008cp} and cellular imaging~\cite{Chang:2010uh}. Most of these studies rely on the properties of isolated single centers, but some applications can benefit from a high density of NV$^-$ centers while keeping a long spin coherence ~\cite{Stanwix:2010ko}, e.g. the realization of collective quantum memories~\cite{Imamoglu:2009eg}, hybrid quantum circuits in which superconducting qubits are coupled to NV$^-$ electron spins~\cite{Kubo:2010iq}, as well as ultrasensitive magnetometry at the micrometer scale~\cite{Acosta:2010dw,Taylor:2008cp}. Furthermore, the application of diamond nanoparticles as labels for bio-imaging also requires a high NV$^-$ content~\cite{Chang:2008ia,Faklaris:2010ba}.
A well-known route to produce a high concentration of NV$^-$ centers is the irradiation by high-energy particles and the subsequent annealing of nitrogen-rich diamond. 

In this work we determined the optimal parameters for the production of such a high NV$^-$ concentration. Several type~1b diamond samples were irradiated with 2.4~MeV protons at fluences ranging from 10$^{12}$ to 10$^{17}$~cm$^{-2}$, and annealed under vacuum at temperatures from 600$^\circ$C to 1000$^\circ$C for durations from 1 to 20 hours. Each sample was characterized at different stages of the NV$^-$ formation process (before/after irradiation and after annealing) using photoluminescence (PL) spectroscopy and slow positron beam based Doppler-broadening spectrometry,  referred in short as Positron annihilation spectroscopy (PAS). The latter technique provides information about the nature of the vacancy defects produced along the depth probed by the positrons~\cite{KrauseRehberg:vl,Note1}.

\section{Experimental Procedure}
The samples used are one-side polished $3\times3\times0.5$~mm$^3$ type 1b HPHT diamond plates (Element Six Ltd.) with a nitrogen concentration specified below 200~ppm ($3.5\times10^{19}$~cm$^{-3}$). Sample irradiation was performed with a 2.4~MeV proton scanning beam from a Van de Graff accelerator. During irradiation, the temperature of samples was maintained below 80$^\circ$C using a water-cooled sample holder. SRIM simulations\footnote{See supplementary material online.} show that such protons create an almost constant concentration of atomic displacements in the first 25~$\mu$m beneath the surface and stop at about 35~$\mu$m producing cascades, which increase sharply the local damage concentration. The sample annealing was performed under vacuum. 
PL measurements were done with a confocal microscope equipped with an imaging spectrograph. We estimate the NV$^-$ concentration by quantitative comparison of the emission spectrum to the one of a single color center\cite{Note1}. 

For PAS measurements we used an accelerator delivering monoenergetic positrons with energy varying from 0.1 to 25~keV (for a complete description see Ref.~\cite{Desgardin:2001uu}). A Doppler broadening spectrometer coupled to the positron accelerator allows us to measure the energy of the gamma photon emitted from the electron-positron annihilation and hence to probe the vacancy defects. 
We determined the usual PAS parameters: $S$ (resp. $W$) corresponds to the annihilation fraction of low (respectively high) momentum electron-positron pairs, probing predominantly valence (respectively core) electrons~\cite{Note1}. $S$ and $W$ change with an increase in the vacancy defect concentration or with a change in the nature of the defects probed. The measurements of $S(E)$ and $W(E)$ as a function of the positron energy $E$ allow to probe the sample as a function of the depth in the first 5~$\mu$m under the surface.
\section{Results and discussions}
PAS and PL measurements were first performed on untreated as-received diamond samples, then on irradiated but not annealed samples, and finally on irradiated and annealed ones.

For all PAS measurements presented in this work, the positron annihilation characteristics $S(E)$ and $W(E)$ as a function of the positron incident energy $E$, can be viewed as resulting from a combination of surface and bulk characteristics. $S(E)$ and $W(E)$ exhibit plateaus at high positron energy (12-25~keV) (see Fig.~\ref{fig1}(b)-(c) and inset of Fig.~\ref{fig2}(a)), indicating that the contribution of the surface can be considered as negligible at the corresponding depths $>1~\mu$m and hence $(S,W) $ values are solely those of the bulk. From now on, we shall only refer to these plateau values and we consider that the distribution of vacancies is homogeneous in the region probed by the positrons.

\subsection{Photoluminescence and positron annihilation measurements done on as-received samples}
\begin{figure}
 \includegraphics[scale=0.50]{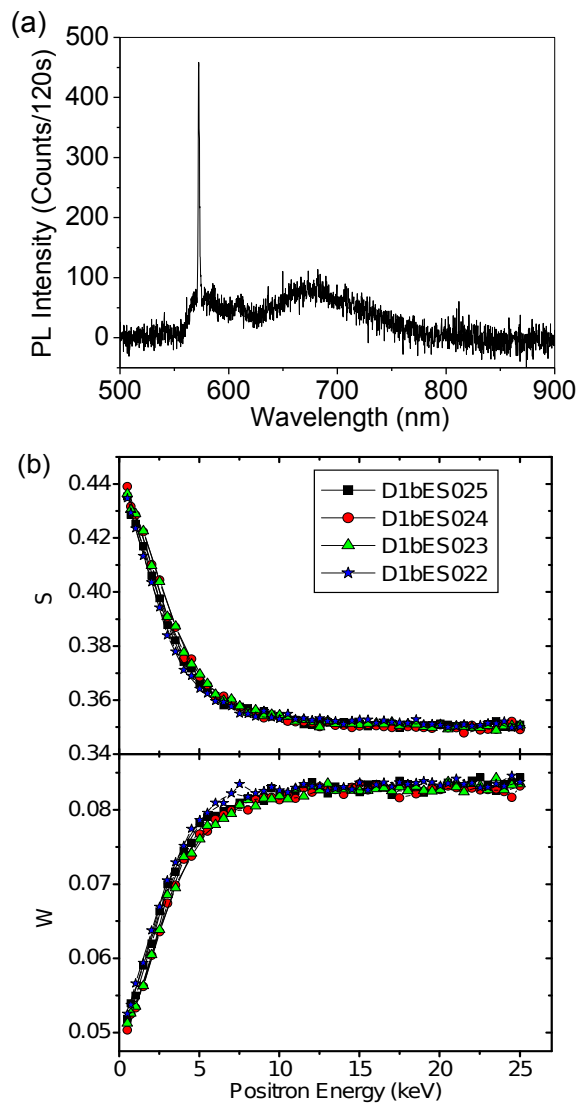}
 \caption{(a) PL of as-received diamond sample (cw laser excitation at the wavelength of 532~nm, laser power: 13.5~$\mu$W), (b) Low momentum annihilation fraction $S$ and high momentum annihilation fraction $W$ of positrons as a function of their incident energy $E$, in four as-received diamond samples identified by references D1bES022-25.
 \label{fig1}}
 \end{figure}
Figure~\ref{fig1}(a) shows that the as-received sample exhibits a sharp PL peak at around 573~nm corresponding to the first order Raman peak of pure diamond at the excitation wavelength of 532~nm. A broader peak around 690~nm indicates the presence of native NV$^-$ color centers. A concentration of $8.3\times10^{12}$~cm$^{-3}$ of native NV$^-$ is estimated from the integration of this peak~\cite{Note1}.

PAS measurements conducted on four different untreated samples show similar $S(E)$ and $W(E)$ graphs (Fig.~\ref{fig1}(b)-(c)), indicating the homogeneity of diamond quality in the set of samples with respect to the atomic structures to which the positrons are sensitive.
From these measurements, we inferred: $S_{\rm ref}=0.351\pm0.001$ and $W_{\rm ref}=0.0834\pm0.0006$.

A fit of the data using VEPFIT~\cite{Note1} allows us to deduce a positron effective diffusion length, $\ell^+$, of about 150~nm. This value is comparable to the value of 111~nm, which was found for $\ell^+$ in type 1b diamond~\cite{Fischer:1999tw}. This shows that we can consider the bulk annihilation characteristics $(S_{\rm ref}, W_{\rm ref})$ measured in the as-received samples as those of the pure diamond lattice $(S_{\rm L}, W_{\rm L})$. This implies that the concentration of vacancy-related defects present in the untreated type 1b samples is below the sensitivity threshold of the PAS technique corresponding to a value of about $10^{15}$~cm$^{-3}$~[Fig.~\ref{fig5}(b)].

Moreover, in the following, we will only consider the relative values $S/S_{\rm L}$ and $W/W_{\rm L}$ which are less sensitive to the experimental setup.

\subsection{Photoluminescence and positron annihilation measurements done after irradiation and prior to thermal annealing}
 \begin{figure}
 \includegraphics[scale=0.65]{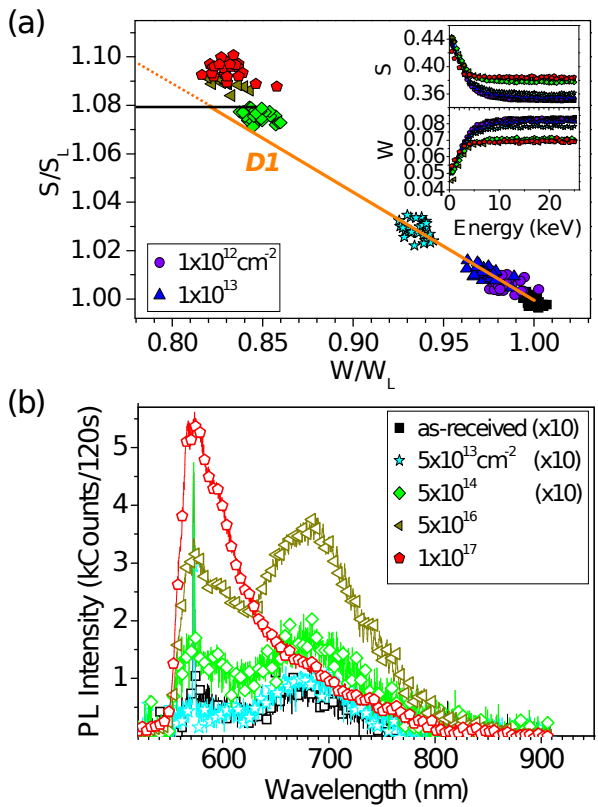}
 \caption{PAS and PL measurements realized on diamond samples after proton irradiation (fluences from $10^{12}$ to $10^{17}$~cm$^{-2}$) and before annealing. (a)$(S/S_{\rm L},W/W_{\rm L})$ scattered plot, with $S$ and $W$ taken at their plateau values in the 12-25~keV energy range (min of 10 measurements per fluence), $S_{\rm L}$ and $W_{\rm L}$ being the diamond lattice parameters\cite{Note1}; inset: $S$ and $W$ as a function of the positron energy (0.1-25~keV). (b) PL spectra measured on the same samples. Excitation with a cw laser (wavelength: 532~nm,  power: 13.5~$\mu$W). The narrow peak at 573~nm is the first order Raman scattering.
 \label{fig2}}
 \end{figure}
\subsubsection{Photoluminescence measurements and inferred NV$^-$ color centers concentration}
PL measurements done on samples irradiated at different proton fluences [Fig.~\ref{fig2}(b)] display a broadband around 690~nm due to NV$^-$ photoluminescence and a higher energy PL band cut below 550~nm by the dichroic beamsplitter. This band is barely visible at the lowest fluence but becomes predominant at the highest fluences. It is tentatively attributed to the H3 (N-V-N) center emission~\cite{Zaitsev:2001tx}. 
The absence of the GR1 center emission (around 741~nm) associated to the neutral monovacancy is noteworthy. It has been shown that in type 1b diamond the monovacancy is preferentially in the negative charge state~\cite{Davies:1992hj} due to the presence of the nitrogen impurity, which can donate an electron to the monovacancy. Both charge states coexist when there are not enough nitrogen donors with respect to vacancies. For the fluence $5\times10^{16}$~cm$^{-2}$ the maximal vacancy concentration can be estimated by that of the atomic displacements calculated by SRIM~\cite{Note1} to be $5\times10^{19}$~cm$^{-3}$, which is larger than the specified nitrogen concentration $3.5\times10^{19}$~cm$^{-3}$. The absence of the GR1 line therefore indicates that the real monovacancy concentration is lower than the one of atomic displacements, as a result of vacancy-interstitial recombination or  non-photoluminescent complex vacancy formation.

\subsubsection{Positron annihilation measurements}
Figure~\ref{fig2} (a) shows that when we increase the fluence, $S/S_{\rm L}$ increases and $W/W_{\rm L}$ decreases indicating the formation of vacancy defects. For fluences $\le5\times10^{13}$~cm$^{-2}$, the data are well fitted by a straight line D1 starting from the as-received sample measurements. This line is the signature of an annihilation state additional to the lattice one and consecutive to irradiation. Since the annihilation fraction in this state increases with the fluence, we suggest that it corresponds to the negatively charged monovacancy, which is the most simple defect that can be created in this type of nitrogen-rich diamond~\cite{Pu:2000cg}.

For  fluences $\ge 5\times10^{14}$~cm$^{-2}$ the $(S, W)$ points are located above D1, indicating that a new type of vacancy-related defect appears~\cite{Liszkay:1994dd}. This claim is supported by the fact that we measure larger relative changes of $S$ values than the one of the neutral or negative monovacancy in diamond~\cite{Pu:2000cg} ($S_{\rm V^\circ}/S_{\rm L}=1.08\pm0.01$ and $S_{\rm V^-}/S_{\rm L}=1.079\pm0.002$) indicated by the horizontal black line in Fig.~\ref{fig2} (a)\footnote{Interestingly, we can also estimate $W_{\rm V^-}/W_{\rm L}=0.82\pm 0.05$} from the intersection of this monovacancy horizontal line with D1. In order to check if NV$^-$ corresponds to this defect center, we estimated its highest concentration from the PL spectrum  [Fig.~\ref{fig2}(b)], to be $1.5\times10^{15}$~cm$^{-3}$ at fluence $5\times10^{16}$~cm$^{-2}$. Such a NV$^-$ concentration is hardly detectable by PAS as proved by further measurements on annealed samples [Fig.~\ref{fig5}(b)]. Moreover this concentration is over-estimated due to the contribution of the high-energy band in the PL spectrum. Therefore NV$^-$ centers present in the samples after irradiation but prior to annealing can only marginally contribute to the $(S/S_{\rm L}, W/W_{\rm L})$ shift. This deviation to the D1 observed at the highest fluences is most likely related to multiple vacancy defects.
 \begin{figure}
 \includegraphics[scale=0.35]{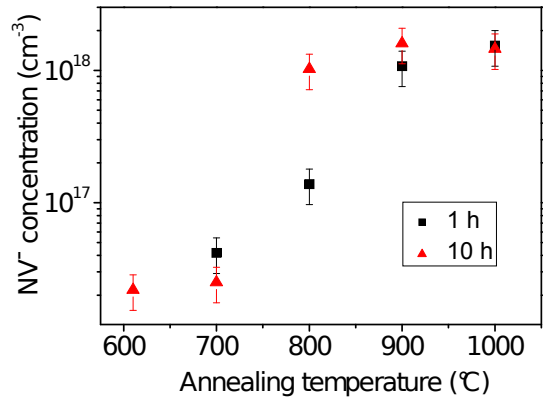}
 \caption{Concentration of NV$^-$ in different diamond samples irradiated at $5\times10^{16}$~cm$^{-2}$ as a function of the annealing temperature (600-1000$^\circ$C) and duration (1 or 10~hours).
 \label{fig3}}
 \end{figure}

\subsection{Post-thermal annealing studies}
\subsubsection{NV$^-$ center formation at fixed proton fluence of $5\times10^{16}$~cm$^{-2}$ and various annealing conditions}
To investigate the effect of annealing time and temperature, diamond samples were irradiated at a proton fluence of $5\times10^{16}$~cm$^{-2}$ and annealed at temperatures ranging from 600$^\circ$C to 1000$^\circ$C, during 1 or 10~hours. Figure~\ref{fig3} displays the estimated NV$^-$ concentration for the different conditions. A significant increase in the NV$^-$ concentration above 700$^\circ$C is observed. Furthermore [NV$^-$] saturates at $\simeq 1.7\times10^{18}$~cm$^{-3}$.
Note that from 800$^\circ$C to 1000$^\circ$C, the difference in NV$^-$ concentration for the two annealing times becomes smaller, eventually vanishing at 1000$^\circ$C. The NV$^-$ formation kinetics is thus enhanced upon temperature increase. 
 \begin{figure}
 \includegraphics[scale=0.55]{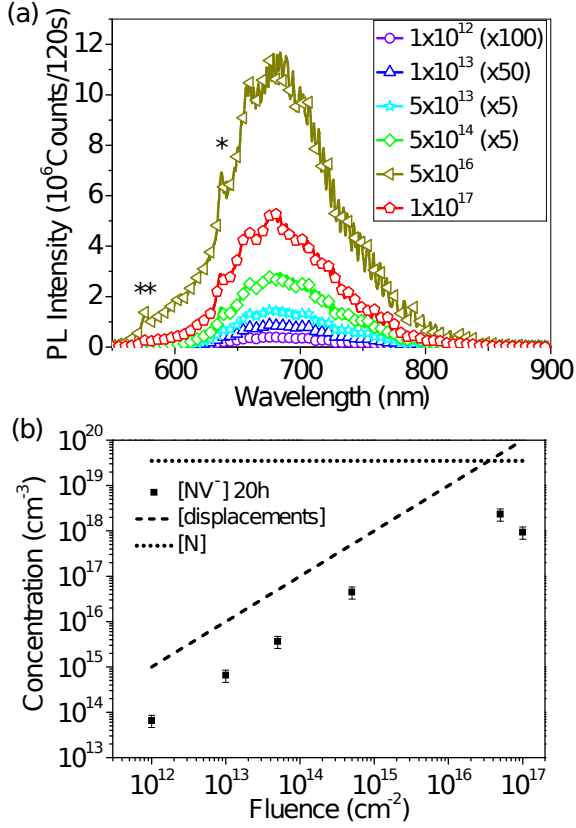}
 \caption{PL  spectra (a) and inferred NV$^-$ concentration (b) in diamond samples as a function of the proton fluence, under identical 800$^\circ$C and 20 h annealing conditions. At high proton fluence, nitrogen converts also into the non-paramagnetic neutral NV$^\circ$ center~\cite{Waldermann:2007hh} characterized by a zero-phonon line (marked by **) at 575~nm wavelength, while the one characteristic of the NV$^-$ is at 637~nm (*). In graph (b) the dashed (dotted) line displays the concentration of atomic displacements predicted by SRIM (resp.  [N]=200~ppm).
 \label{fig4}}
 \end{figure}
 \subsubsection{NV$^-$ center formation for fixed annealing conditions  (800$^\circ$C, 20 hours) and various fluences}
We also studied the influence of the irradiation fluences (from $10^{12}$~cm$^{-2}$ to $10^{17}$~cm$^{-2}$) on NV$^-$ center formation for a given annealing condition (800$^\circ$C, 20 hours). The PL spectra of Fig.\ref{fig4}(a) show an intensity increase with the fluence up to $5\times10^{16}$~cm$^{-2}$. Yet at all the fluences, the inferred NV$^-$ concentration plotted on Fig.~\ref{fig4}(b) is one order of magnitude lower than that of the atomic displacements predicted by SRIM, probably due to vacancy-interstitial recombination. The maximum NV$^-$ concentration reached is $2.3\times10^{18}$~cm$^{-3}$. Note that a lower NV$^-$ concentration is measured for the highest fluence. This decrease could have been attributed to the onset of diamond amorphization but confocal Raman spectroscopy did not reveal any sp2 carbon signal~\cite{Note1}. However, PAS measurements provides an explanation as shown in the next paragraph.

 \begin{figure}
 \includegraphics[scale=0.60]{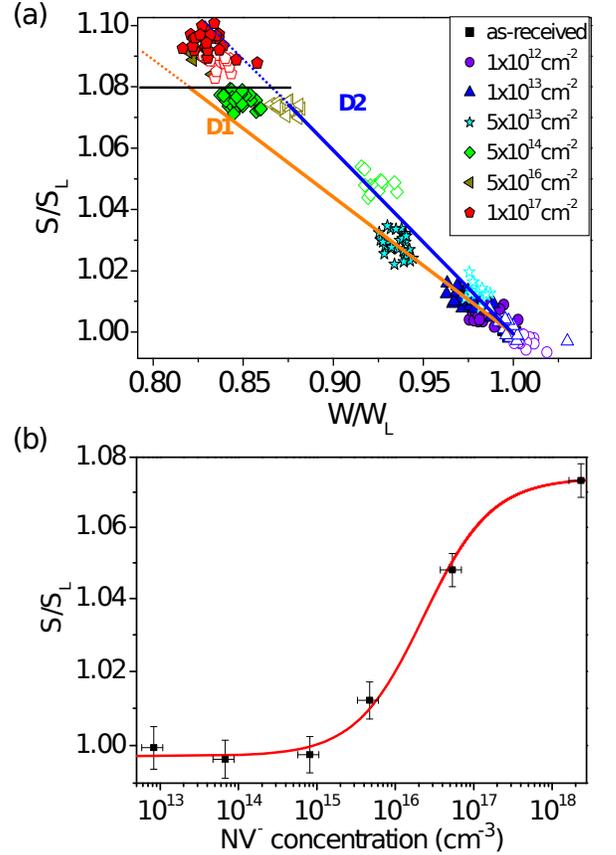}
 \caption{PAS measurements on samples irradiated at different fluences and annealed at 800$^\circ$C for 20~h. (a) $(S/S_{\rm L}, W/W_{\rm L})$ scattered plots. Filled (open) symbols refer to irradiated but non-annealed (resp. annealed) samples. Black squares refer to the untreated sample. (b) $S/S_{\rm L}$ vs. NV$^-$ concentration in annealed samples; plain curve: best fit by relation~(\ref{eq1}).
 \label{fig5}}
 \end{figure}
Figure~\ref{fig5}(a) displays $S(W)$ for the different fluences before and after annealing. At all the fluences, we observe a decrease of $S$ concomitant with an increase of $W$ after annealing, indicating that the vacancy defect concentration diminishes and/or the nature of the defects changes. $(S/S_{\rm L}, W/W_{\rm L})$ points after annealing are aligned on a new straight line D2 except the highest fluence one. Since PL measurements evidence the creation of NV$^-$ defect, we propose that it is the defect associated to D2. This claim is supported by previous work showing an increase of $W$ due to nitrogen decoration of vacancies in diamond~\cite{SACHDEVA:2004hf}. Note that for $5\times10^{14}$~cm$^{-2}$ and $5\times10^{16}$~cm$^{-2}$ fluences $(S/S_{\rm L}, W/W_{\rm L})$ points which are initially off the monovacancy D1 line finally end onto D2 after annealing.
It could be due to the positron trapping rate balance between different vacancy traps shifting the $S$ and $W$ values to the annihilation characteristics of the NV$^-$ vacancy defect which has a high trapping coefficient, without excluding that annealing can partially remove the complex vacancies. 
In contrast, for the highest fluence the $(S/S_{\rm L}, W/W_{\rm L})$ points almost did not move after annealing, and still fall off the D2 line, which indicates that there is a significant fraction of complex vacancies that do not evolve into NV$^-$ centers, thus limiting the NV$^-$ creation yield.
It would be worthwhile to anneal at temperatures higher than 1000$^\circ$C. This strategy was successful in increasing the NV$^-$ electronic spin coherence as a result of paramagnetic defect destruction~\cite{Naydenov:2010ks} including divacancies associated ones like R4 or W6~\cite{Twitchen:1999ed}.

\subsubsection{Trapping coefficient of positrons by NV$^-$ suggested by positron annihilation and photoluminescence measurements}
The combination of PAS and PL measurements also allows us to estimate the trapping coefficient of positrons by NV$^-$. Fig.~\ref{fig5}(b) displays $S$ as a function of the NV$^-$ concentration inferred from PL spectra [Fig.~\ref{fig4}(a)]. The data can be fitted by a relationship derived from rate equations~\cite{Pu:2000cg} considering the lattice and vacancy defects (mostly NV$^-$) as the positron trapping sites:
\begin{equation}
\frac{S}{S_{\rm L}}=\frac{\lambda_{\rm L}}{\lambda_{\rm L}+\mu_{\rm NV^-}C/C_{\rm L}}+\frac{\mu_{\rm NV^-}C/C_{\rm L}}{\lambda_{\rm L}+\mu_{\rm NV^-}C/C_{\rm L}}\frac{S_{\rm NV^-}}{S_{\rm L}},
\label{eq1}
\end{equation}
where  $\lambda_{\rm L}$ is the lattice annihilation rate, $C$ and $C_{\rm L}=1.8\times10^{23}$~cm$^{-3}$ the NV$^-$ center concentration and diamond lattice atomic density respectively. $\mu_{\rm NV^-}$ is the associated positron trapping rate. Taking $\lambda_{\rm L}=10$~ns$^{-1}$ from Ref.~\cite{Pu:2000cg}, the best fit to the data is obtained for $S_{\rm NV^-}/S_{\rm L}=1.074\pm 0.005$ and $\mu_{\rm NV^-}=(7.6\pm1.6)\times10^{16}~$s$^{-1}$.
The same type of fit can be performed for $W/W_{\rm L}$, yielding $W_{\rm NV^-}/W_{\rm L}=0.82\pm 0.05$ and $\mu_{\rm NV^-}=(6.7\pm 1.8)\times10^{16}$~s$^{-1}$. The two values of $\mu_{\rm NV^-}$ obtained are consistent, and we take their mean value $\mu_{\rm NV^-} =(7.2\pm1.8)\times10^{16}~$s$^{-1}$ as the one inferred from PAS. In silicon the positron trapping coefficient has been found to be  $(0.8\pm0.4)\times10^{15}$~s$^{-1}$ for the neutral divacancy~\cite{Mascher:1989bh} and $1.8\times10^{16}~$s$^{-1}$ for the (VP)$^-$ defect~\cite{Kawasuso:1995ez}. The more efficient trapping of V$^-$ in diamond compared to (VP)$^-$ in silicon can be explained by: (1) a smaller static dielectric constant (5.6 for diamond and 11.6 for silicon), hence a longer range coulomb potential in the case of diamond, (2) a binding energy of 0.3~eV for the Rydberg state of the V$^-$ electron, larger than the one of 0.1~eV in silicon, leading to a significant reduction of the positron detrapping in diamond. Table~\ref{table1} summarizes the PAS data related to the defects detected in this work.
\begin{table}
 \caption{Positron annihilation parameters and trapping rate $\mu$ for the vacancy defects in diamond, as estimated in this work (except for $S_{V^-}/S_{\rm L}$).\label{table1}}
 \begin{ruledtabular}
 \begin{tabular}{lll}
Positron parameters							& V$^-$ 							&	NV$^-$ \\
\hline
$S/S_{\rm L}\;(S_{\rm L}=0.351\pm0.001)$		& $1.079\pm 0.002$~\cite{Pu:2000cg} 	& 	$1.074\pm 0.005$ \\
$W/W_{\rm L}¥(W_{\rm L}=0.0834\pm0.0006)$		& $0.82\pm 0.05$					& 	$0.87\pm0.01$ \\
$\mu\; (\times10^{16}~$s$^{-1})$					& -								&	$7.2\pm1.8$ \\
 \end{tabular}
 \end{ruledtabular}
\end{table}
\section{Conclusion}
We have optimized the formation of a high-density of NV$^-$ centers in type 1b diamond by proton irradiation and subsequent thermal annealing. We showed that the conversion efficiency of nitrogen into NV$^-$ increases with the irradiation fluence, up to a point where further increase leads to a reduction in this conversion rate, most likely due to the creation of complex vacancies stable under 800$^\circ$/20~h annealing. We achieved the maximum NV$^-$ concentration of $2.3\times10^{18}$~cm$^{-3}$ (13.5~ppm), corresponding to a conversion efficiency [NV$^-$]/[N] of about 6.7\% ([N]$\simeq200$~ppm).  The positron trapping coefficient by NV$^-$ centers in diamond was estimated for the first time.

\section*{Acknowledgement}
We thank J.-F.~Roch, V.~Jacques and A.~Trifonov for fruitful discussions, and M.-R.~Ammar for help in Raman measurements. This work was supported by the French National Research Agency (ANR-07-NANO-045).

\bibliography{ref}

\end{document}